\documentclass[aps,prl,twocolumn,superscriptaddress,showpacs]{revtex4}
\usepackage{graphics}
\usepackage{graphicx}
\usepackage[english]{babel}
\usepackage{color}

\begin{document}

\title{Enhancement of penetration field in vortex nanocrystals due to Andreev bound states}

\author{M. I. Dolz}
\affiliation{Departamento de F\'{i}sica, Universidad Nacional de
San Luis and CONICET,  San Luis, Argentina}

\author{N. R. Cejas Bolecek}
\affiliation{Centro At\'{o}mico Bariloche and Instituto Balseiro,
CNEA, CONICET and Universidad Nacional de Cuyo, Bariloche,
Argentina}

\author{J. Puig}
\affiliation{Centro At\'{o}mico Bariloche and Instituto Balseiro,
CNEA, CONICET and Universidad Nacional de Cuyo, Bariloche,
Argentina}

\author{H. Pastoriza}
\affiliation{Centro At\'{o}mico Bariloche and Instituto Balseiro,
CNEA, CONICET and Universidad Nacional de Cuyo, Bariloche,
Argentina}

\author{G. Nieva}
\affiliation{Centro At\'{o}mico Bariloche and Instituto Balseiro,
CNEA, CONICET and Universidad Nacional de Cuyo, Bariloche,
Argentina}

\author{J. Guimpel}
\affiliation{Centro At\'{o}mico Bariloche and Instituto Balseiro,
CNEA, CONICET and Universidad Nacional de Cuyo, Bariloche,
Argentina}

\author{C.J. van der Beek}
\affiliation{Centre de Nanosciences et de Nanotechnologies, CNRS,
Universit\'e Paris-Sud, Universit\'e Paris Saclay, Palaiseau,
France}

\author{M. Konczykowski}
\affiliation{Laboratoire des Solides Irradi\'{e}s, CEA/DRF/IRAMIS,
Ecole Polytechnique, CNRS, Institut Polytechnique de Paris,
Palaiseau, France}

\author{Y. Fasano}
\affiliation{Centro At\'{o}mico Bariloche and Instituto Balseiro,
CNEA, CONICET and Universidad Nacional de Cuyo, Bariloche,
Argentina}

\date{\today}

\begin{abstract}
We study the penetration field $H_{\rm P}$ for vortex nanocrystals
nucleated in micron-sized samples with edges aligned along the nodal
and anti-nodal directions of the d-wave superconducting parameter of
Bi$_2$Sr$_2$CaCu$_2$O$_{8 - \delta}$.  Here we present evidence that
the $H_{\rm P}$ for nanocrystals nucleated in samples with edges
parallel to the nodal direction is  larger than for the antinodal
case, $\sim 72$\,\% at low temperatures. This finding supports the
theoretical proposal that surface Andreev bound states appearing in
a sample with edges parallel to the nodal direction would produce an
anomalous Meissner current that increases the Bean-Livingston
barrier for vortex penetration.This has been detected thanks to the
nucleation of vortex nanocrystals with a significant
surface-to-volume ratio.
\end{abstract}

 \keywords{}

\maketitle

The current need for miniaturization of devices calls for a better
understanding on how the thermodynamic, structural and magnetic
properties of condensed matter are affected by decreasing the sample
size. The components of these devices are made of functional
materials as small as nanocrystals with hundred particles or less.
The physical properties of nanocrystals of hard condensed matter
have been studied both experimentally and theoretically for several
systems as metallic nanoparticles, semiconductor nanocrystals and
films.~\cite{Coombes1972,Goldstein1992,Tolbert1994,Guisbiers2009}
Some thermodynamic properties of phase transitions, such as the
transition temperature, the entropy and enthalpy jumps, are depleted
when reducing the size of components at the
nanoscale.~\cite{Coombes1972,Goldstein1992,Tolbert1994,Guisbiers2009}
This  is ascribed to a reduction in the average binding energy of
the nanocrystal due to the high proportion of particles located at
the surface having a lesser binding energy than those of the volume.

The nucleation of nanocrystalline vortex matter in micron-sized
superconducting samples opens the possibility of studying this
general problem in soft condensed matter
systems.~\cite{Moshchalkov1995,Geim1997,Schweigert1998,Wang2001,Dolz2015}
Surface effects in vortex nanocrystals affect its thermodynamic,
structural and magnetic
properties.~\cite{Palacios1998,Dolz2014a,CejasBolecek2015,CejasBolecek2017}
For instance, due to surface barriers, the field at which the first
vortex penetrates, $H_{\rm P}$,~\cite{Konczykowski1991,Clem2008} can
be larger than the lower critical field $H_{\rm c1}$ (depending only
on the penetration depth $\lambda$ and the coherence length $\xi$ of
the material). This effect becomes more relevant when the
surface-to-volume ratio of the number of vortices in nanocrystals
enhances.~\cite{Wang2001,Dolz2014a} Surface barriers produce
hysteretic behavior in the vortex magnetic response and can be of
two types. Geometrical barriers are caused by the extra energy cost
for flux entry produced by the local enhancement, close to the
sample edge, of the otherwise uniform outer  field
$H$.~\cite{Zeldov1994,Willa2014} Bean-Livingston (BL) barriers arise
from the competition between the Meissner current pushing the vortex
inside the sample versus the attraction of the vortex towards the
outer image-vortex.~\cite{Koshelev1992,Burlachkov1993} On increasing
H, the former term dominates and vortices penetrate the sample.

Theoretical studies predict that nucleating vortices in anisotropic
d-wave superconductors with Andreev bound states (ABS) can modify
the BL surface barrier, and therefore $H_{\rm
P}$.~\cite{Fogelstrom1997}
 ABS are surface states
localized within a distance $\sim \xi$ from the edge of the
samples,~\cite{Hu1994} of a few nanometers for high-$T_{\rm c}$'s.
These zero-energy excitations appear if the sample edge is oriented
along the nodal direction of the d-wave superconducting parameter
and generate an anomalous current running opposite to the
supercurrents, namely $\mathbf{J_{SC}} = 2e \mid \Psi \mid^{2}(2e/cm
(\mathbf{A_{0}} - \mathbf{A'}))$.~\cite{Iniotakis2005} The first
term is the regular Meissner current and the second is  due to the
vector potential of the eventually stable ABS.~\cite{Iniotakis2008}
Therefore, a dependence of $H_{\rm P}$ on the orientation of the
sample edge with the anisotropic d-wave order parameter is
expected.~\cite{Iniotakis2008} The decrease in $J_{SC}$ for samples
with edges parallel to the nodal direction produces an enhancement
of the BL barrier: the term pushing the vortex inside has a lesser
magnitude but the attraction towards the outer anti-vortex remains
unaltered.

These surface effects are expected to emerge in the magnetic
properties of superconducting samples when dramatically decreasing
the size of the vortex crystal. In this work we study vortex
nanocrystals with 10$^3$-10$^6$ vortices and 10 to 1\,\%
surface-to-volume ratio nucleated in micron-sized
Bi$_2$Sr$_2$CaCu$_2$O$_{8 - \delta}$ cuboids. We study two sets of
cuboids, with edges aligned along the nodal (N) and the anti-nodal
(AN) directions of the d-wave order parameter.  We found that
$H_{\rm P}$ is enhanced in N  with respect to AN cuboids of the same
size, up to $\sim 72$\,\% at low temperatures. This result has been
detected thanks to the virtuous combination of low-noise local
magnetic techniques and the nucleation of vortex nanocrystals.

\begin{figure}[ttt]
\includegraphics[width=0.9\columnwidth,angle=0]{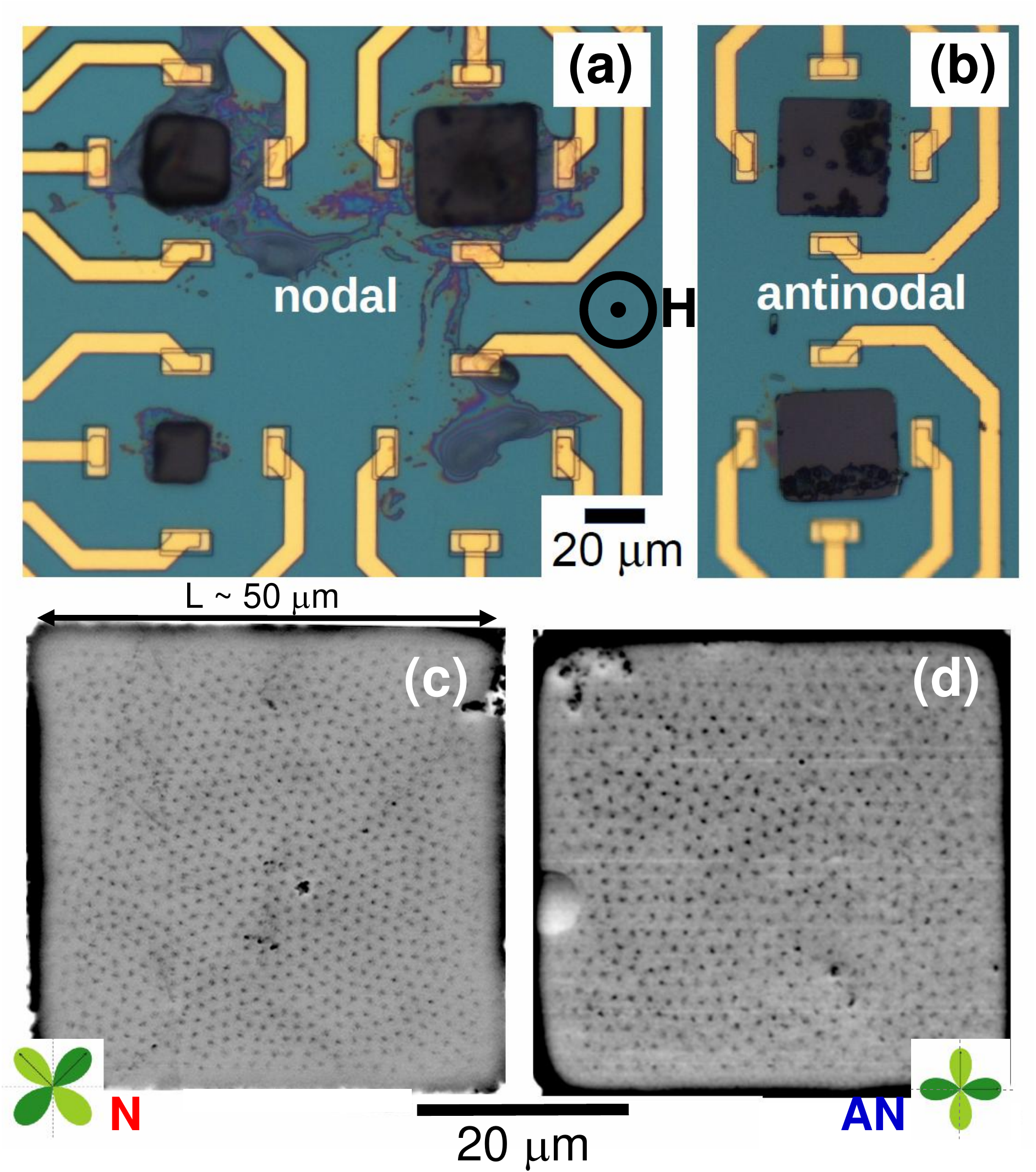}
\caption{Some of the nodal (N) and antinodal (AN)
Bi$_{2}$Sr$_{2}$CaCu$_{2}$O$_{8 + \delta}$ cuboids studied located
on top of $16 \times 16$\,$\mu$m$^{2}$ Hall sensors. (a) N cuboids
with sides $L=20$, 30 and 50\,$\mu$m; (b) AN cuboids with $L \sim
50$\,$\mu$m. Vortex nanocrystals imaged by magnetic decoration in
(c) N (8.5\,Gauss) and (d) AN (5\,Gauss) cuboids and alignment
 of the sample edge with the nodes
of the d-wave superconducting parameter.\label{figure1}}
\end{figure}

We micro-engineer the cuboids from optimally-doped
Bi$_{2}$Sr$_{2}$CaCu$_{2}$O$_{8}$ macroscopic
single-crystals~\cite{Li94a} with critical temperature $T_{\rm
c}=90$\,K. We start from rectangular macroscopic samples aligned
with their edges parallel to the $a$ (or $b$) crystalline direction,
i.e. parallel to the N direction of the d-wave order parameter.
Cuboids were fabricated by combining optical lithography and
physical ion milling.~\cite{Dolz2015}. N cuboids are obtained by
aligning the optical-lithography mask such that their sides result
parallel to the  macroscopic sample edge. In a similar fashion, AN
cuboids were fabricated rotating the lithography mask 45\,$^{\circ}$
from the previous configuration.~\cite{Supplemental} We studied
several N and AN cuboids with square sides $L=20$, 30 and
50\,$\mu$m, thicknesses $t\sim 2$\,$\mu$m. The critical temperature
of all studied cuboids is the same than that of the parent
macroscopic crystal, $T_{\rm c}=90$\,K.

 Figure\,\ref{figure1}
shows various samples placed on top of Hall-sensors with $16 \times
16$\,$\mu$m$^2$ working areas micro-fabricated from GaAs/AlGaAs
heterostructures. Samples were micro-manipulated and glued with
Apiezon N grease to improve thermal contact. Local ac and dc
magnetization measurements were performed applying dc, $H$, and
ripple, $h_{\rm ac}\ll H$, magnetic fields parallel to the $c$-axis.
The Hall sensor signal is proportional to the  magnetization of the
cuboids, $H_{\rm s} = (B - H)$. We measure dc and ac hysteresis
loops recording $H_{\rm s}$ and the sample transmittivity $T'$ on
sweeping $H$. $T'$ is obtained applying an $h_{\rm ac}$
 and normalizing the in-phase component of the
first-harmonic signal, $B'$. This magnitude is measured by means of
a digital-signal-processing lock-in technique using the lock-in
reference signal as supply to the coil generating $h_{\rm ac}$. The
normalization $T'=[B'(T) - B'(T \ll T_{\rm c})]/[B'(T>T_{\rm c}) -
B'(T \ll T_{\rm c})]$ is such that $T'=1$  in the normal state and
$T'=0$ well within the superconducting phase.

The superconducting quality of the cuboids was checked by magnetic
decoration experiments. ~\cite{Fasano1999}  Figure\,\ref{figure1}
  (c) and (d) show vortex nanocrystals nucleated at low vortex densities
in N and AN cuboids with $L=50$\,$\mu$m. Black dots correspond to
individual vortices decorated with magnetized Fe nanoparticles
attracted to the cores due to their local field gradient. Regular
vortex structures with an excess of topological defects induced by
confinement~\cite{CejasBolecek2017} are observed for both, N and AN
cuboids.

\begin{figure}[ttt]
\includegraphics[width=0.85\columnwidth,angle=0]{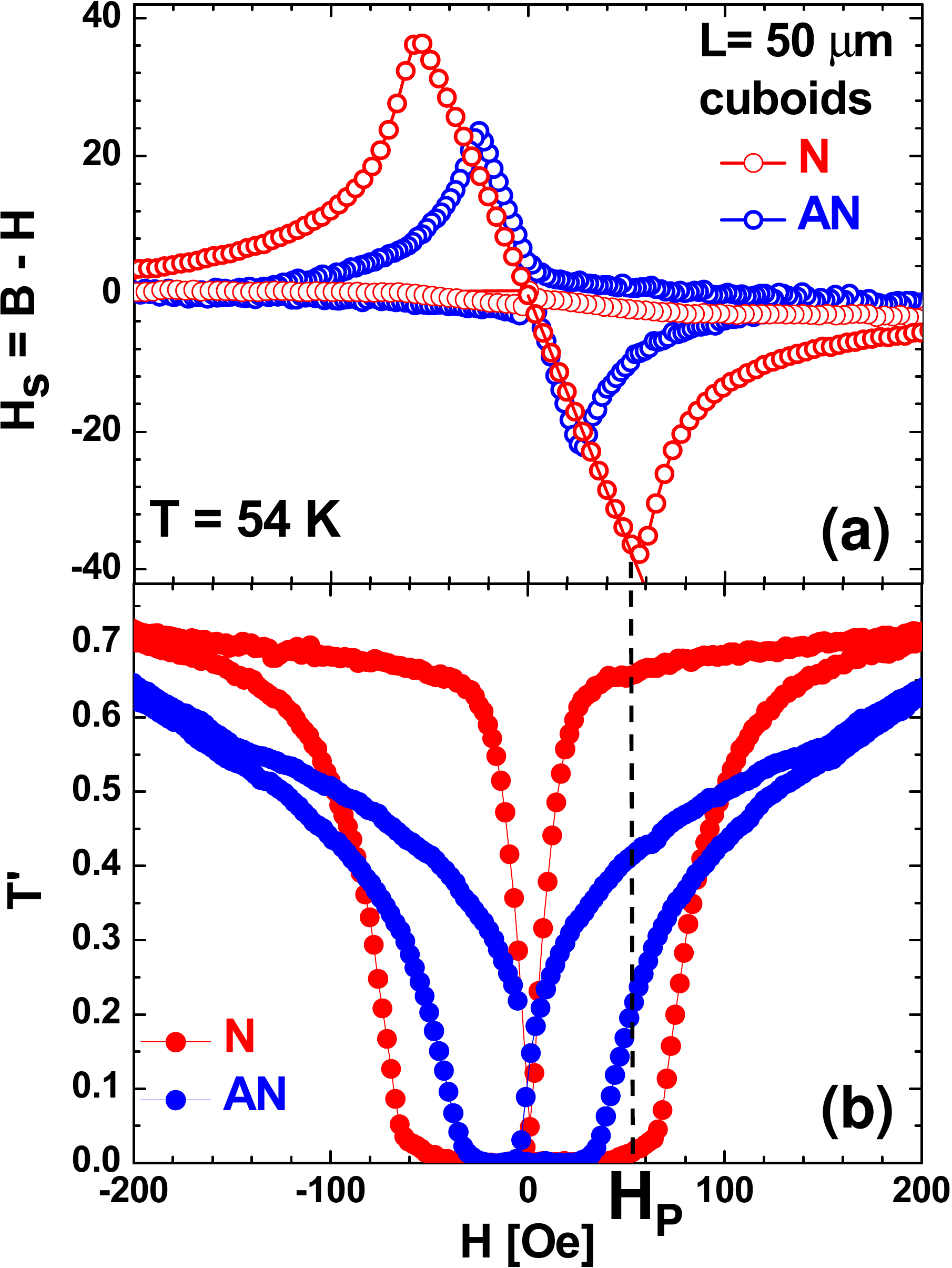}
\caption{Detection of the penetration field $H_{\rm p}$ in nodal (N)
and antinodal (AN) cuboids (with  $L=50$\,$\mu$m) from (a) dc and
(b) ac magnetic hysteresis loops.  The $H_{\rm p}$ determined from
the deviation of the linear Meissner response  coincides with the
field at which transmittivity becomes non-negligible (see vertical
dashed lines). Measurements performed at 54\,K; ac measurements with
$h_{\rm ac}$ of 1\,Gauss and 7.1\,Hz.\label{figure2}}
\end{figure}

Figure\,\ref{figure2} (a) shows illustrative dc hysteresis loops at
54\,K in N and AN cuboids, both with $L=50$\,$\mu$m.  On increasing
field from zero, $H_{\rm s}$ follows first the linear Meissner
response associated to complete field expulsion. The entrance of the
first vortex into the cuboids is signposted by the departure of
$H_{\rm s}$ from linearity, at a field $H_{\rm p}$. On further
increasing field, $H_{\rm s}$ changes curvature as expected when
vortices penetrate. The dc  loops have a two-quadrant locus, with
branches located mostly in the second and forth quadrant. The two
field-descending branches for positive and negative $H$ are almost
horizontal and close to zero. The same behavior is observed for all
dc loops measured between 35 and 90\,K.~\cite{Supplemental} This
phenomenology suggests that for these vortex nanocrystals bulk
pinning has a lesser effect than surface barriers for vortex
penetration.

We also measured $H_{\rm p}$ in N and AN cuboids from ac hysteresis
loops, see Fig.\,\ref{figure2} (b). At low fields $T'=0$, indicating
full expulsion of the magnetic field; on increasing field, $T'$
becomes non-negligible at roughly the same $H_{\rm p}$  where the
departure of linearity is detected in dc loops, see dashed line in
Fig.\,\ref{figure2}. On further increasing $H$ the number of
vortices penetrating the sample enhances and $T'$ grows in
accordance, reaching a value close to 1 for high fields. For a given
$H$ in the same loop, $T'$ in the descending branch is larger than
in the ascending one since in the former there is extra trapped
flux.

\begin{figure}[ttt]
\vspace{-1cm}
\includegraphics[width=0.95\columnwidth,angle=0]{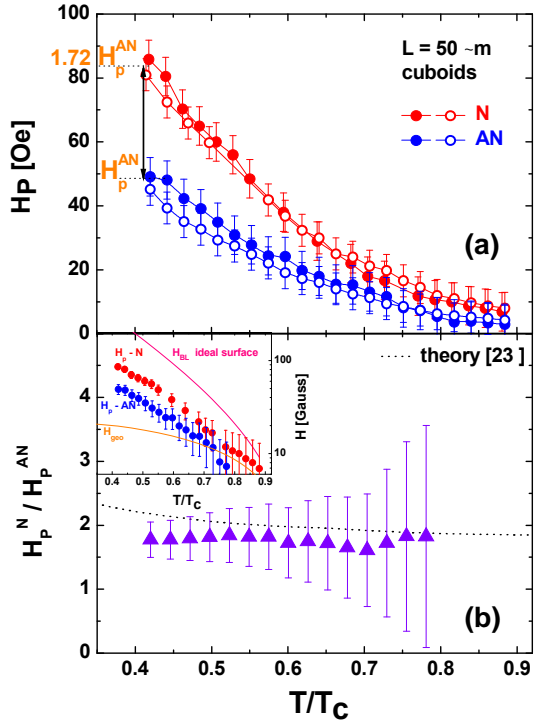}
\caption{(a) Temperature-dependence of the penetration field $H_{\rm
p}$ for nodal (N) and antinodal (AN) cuboids from dc and ac (open
and full symbols) measurements. (b) $H_{\rm p}^{\rm N}/H_{\rm
p}^{\rm AN}$ data (symbols) and temperature-dependence predicted in
Ref.\,\onlinecite{Iniotakis2008} (black dotted line) for a
superconductor with $\kappa=200$. Insert: Data of (a) compared with
the theoretical expectation for geometrical (orange) and BL (pink)
barriers for cuboids of the same size and material than we
study.\label{figure3}}
\end{figure}

The most remarkable result of  dc and ac data of Fig.\,\ref{figure2}
is that the penetration field for N is larger than for AN cuboids.
Figure\,\ref{figure3} (a) shows a comparison between the
temperature-evolution of $H_{\rm p}$ for N and AN cuboids with
$L=50$\,$\mu$m. The values obtained considering the virgin branches
of  ac (open points) and dc (full points) magnetization loops are
similar within the uncertainty.  In all the studied
temperature-range, $H_{\rm p}^{\rm N}$ is larger than $H_{\rm
p}^{\rm AN}$ beyond error bars, with this difference increasing on
cooling. For instance, for the smallest measured temperature,
$T/T_{\rm c} \sim 0.4$, $H_{\rm p}^{\rm N} \simeq 1.72 H_{\rm
p}^{\rm AN}$.

\begin{figure}[ttt]
\includegraphics[width=0.9\columnwidth,angle=0]{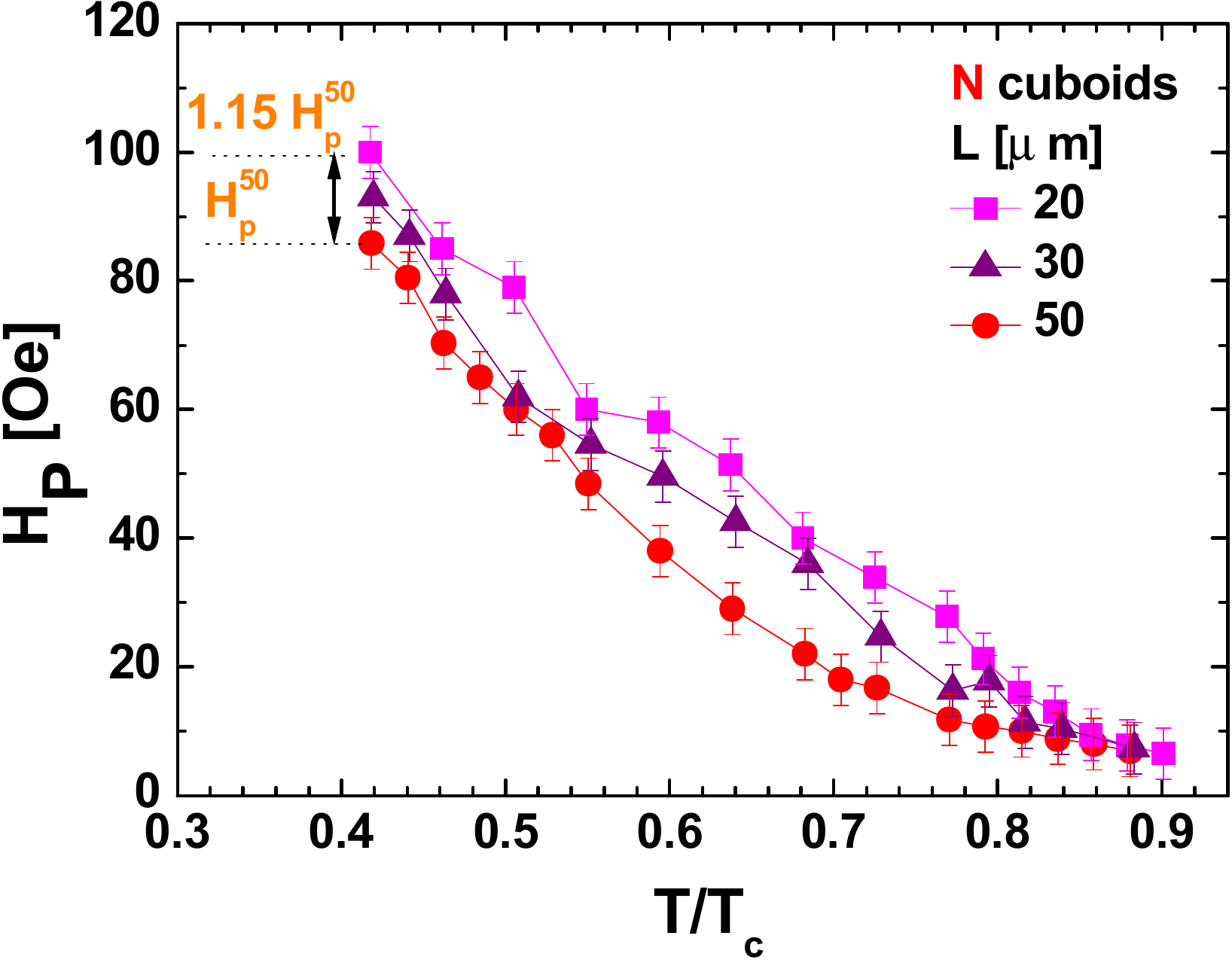}
\caption{Temperature-dependence of the penetration field for nodal
cuboids with different square sides of $L=20$, 30 and 50\,$\mu$m.
Data from ac magnetization in agreement with dc data within the
error. Measurements performed with a ripple field of 1\,Gauss and
7.1\,Hz. \label{figure4}}
\end{figure}

Even though $H_{\rm p}$ is sensitive to the side of the cuboids,
this difference can not be accounted by eventual small changes in
$L$. Figure \,\ref{figure4} shows the temperature-dependence of
$H_{\rm p}$ for N cuboids with $L=20$, 30 and 50\,$\mu$m. The
smaller the $L$, the larger $H_{\rm p}$, but the separation between
curves enlarge on decreasing $T$. In particular, at $T/T_{\rm c}
\sim 0.4$ the $H_{\rm p}$ difference between the 20 and 50\,$\mu$m
cuboids is of $\sim 15$\,\%, much smaller than that found between
$H_{\rm p}^{\rm AN}$ and $H_{\rm p}^{\rm N}$ at the same $T$.

We calculated the ratio $H_{\rm p}^{\rm N}/H_{\rm p}^{\rm AN}$ by
averaging $H_{\rm p}$ data obtained  from ac and dc measurements,
see Fig.\,\ref{figure3} (b). This ratio seems to be featureless in
temperature, even if the error bars are significant.  The figure
also shows the theoretically-expected evolution of $H_{\rm p}^{\rm
N}/H_{\rm p}^{\rm AN}$ when considering the effect of  Andreev bound
states in the BL surface barrier for an ideal infinite sample of a
d-wave superconductor with $\kappa=\lambda/\xi=200$, and neglecting
any other barrier for vortex penetration.~\cite{Iniotakis2008} For
the vortex nanocrystals studied here, $\kappa \sim 200$ as in the
theoretical study, but the cuboids are not an ideal infinite
interface. However, our N and AN experimental data are quite close
to the theoretically-expected BL barrier for an isotropic
superconducting cuboid with $L=50$\,$\mu$m and $t=2$\,$\mu$m, see
pink curve in the insert. This curve was obtained considering that
the temperature-dependence of $H_{\rm p}$ for a BL barrier in an
ideally specular surface is $H_{\rm BL}(T)= H_{\rm
c}(T)\sqrt{t/L}\exp{(-T/T_{\rm 0})}$. $H_{\rm c}(T)\simeq \kappa
H_{\rm c1}(T)/\ln{\kappa}$ is the thermodynamic critical field with
$H_{\rm c1}(T)$ the first critical field, and $T_{\rm 0}\sim 15$\,K
for Bi$_{2}$Sr$_{2}$CaCu$_{2}$O$_{8}$~\cite{Wang2001}, a
characteristic temperature below which pancake vortices are
individually pinned.~\cite{Wang2001,Burlachkov1993} This ideal
$H_{\rm BL}(T)$ coincides with the  data at high $T$, and overpass
them by more than a factor two at low $T$. This discrepancy quite
likely has origin in the non-ideal nature of the edges of our
samples.

Nevertheless, the $H_{\rm p}^{\rm N}/H_{\rm p}^{\rm AN}$
experimental data lay pretty close to the theoretical value expected
when considering the effect of Andreev bound states, even though
this theoretical curve has no other free parameter than the
 $\kappa$ of the superconducting material. In addition,
geometrical barriers do not seem to play a relevant role at
$T/T_{\rm c} \lesssim 0.6$ as to explain the enhancement of $H_{\rm
p}^{\rm N}$ versus that of $H_{\rm p}^{\rm AN}$. Indeed, the insert
to Fig.\,\ref{figure3} (b) shows that $H_{\rm geo}$, the penetration
field associated to the geometrical barrier for the geometry of the
 studied  cuboids is always below the experimental N and AN $H_{\rm
p}$. This curve is obtained considering that $H_{\rm geo}=H_{\rm
c1}(T) \tanh{\sqrt{0.36 t/L}}$ with $0.36$ the geometrical factor
for cuboids.~\cite{Brandt}

In conclusion, the comparison between the theoretical BL and
geometrical barriers in cuboids,  and our experimental data in
cuboids of a d-wave superconductor, suggests the dominance of the BL
barrier for vortex penetration at low $T$. This can be at the origin
of the rather good quantitative agreement between the experimental
data and the theoretically predicted $H_{\rm p}^{\rm N}/H_{\rm
p}^{\rm AN}$ for d-wave ideal infinite superconductors, a striking
result considering that this theoretical calculation depends only in
the value of $\xi$ for the superconducting material. The latter
suggests that the barriers for vortex penetration in N cuboids are
fully governed by the decrease in the effective Meissner current
induced by the presence of Andreev bound states located only up to a
distance $\xi \sim 5\cdot 10^{-4} L$ from the surface of the
cuboids. Therefore, we present here evidence of a new phenomena in
the magnetic properties of vortex nanocrystals emergent from the
microscopic local electronic properties of anisotropic d-wave
superconductors. From an applied point of view, our data suggests
that the crystal orientation of micron-sized superconducting samples
made of high-$T_{\rm c}$'s, with quite frequently anisotropic order
parameters, is a property that has to be taken into account for the
magnetic response of devices based in such tiny building blocks.

We thank V. Mosser for providing the sensors, M. Li for growing the
macroscopic crystals, the ECOS-Sud Argentina-France collaboration
and  grants ANR-10-LABX-0039-PALM and ANR-CE08-007 DiSSCo-Hall  for
financial support.

\end{document}